\documentclass[aps,prl,10pt,twocolumn,groupedaddress]{revtex4-2}
\usepackage{graphicx}
\usepackage[caption=false,labelformat=empty]{subfig}
\usepackage{amsmath,amsfonts,amssymb}
\usepackage{dsfont}
\usepackage{hyperref}
\begin{document}

\title{Circulation of Elites in an Adaptive Network Model}

\author{Alexander Jochim}
\email{jochim@itp.uni-bremen.de}
\affiliation{Institute for Theoretical Physics, University of Bremen, Germany}
\author{Stefan Bornholdt}
\email{bornholdt@itp.uni-bremen.de}
\affiliation{Institute for Theoretical Physics, University of Bremen, Germany}

\begin{abstract}
Societies experience politically stable and unstable phases along history, whereas political power is usually passed to new elite groups by these changes. Structural dynamics of the elites in a society have been proposed to be one of the core drivers shaping long term behavior. As current models and data are rather macroscopic, the emergence of macroscopic behavior from microscopic dynamics is largely unclear. Here, we introduce an adaptive network model of directed links representing political power and competing political ideas, based on local dynamical rules, only. The model is based on two socially motivated behaviors: the cumulative advantage effect of political power and intra-elite conflict. We observe punctuated equilibria as an emergent behavior and find a phase transition towards a disordered phase. We define an advance warning measure for elite collapse and find that the states of only a few largest nodes are suitable as a proxy with predictive information.
\end{abstract}

\maketitle
\section{Introduction}
The concept of elite circulation was introduced by Pareto about 100 years ago \cite{pareto1935mind}, which states that the governing elite groups in a society are continuously replaced as a process of societal growth and renewal. Recent literature picks up this idea and discusses the role of elites as a fundamental driver of recurring societal instability on the basis of empirical data \cite{goldstone1991revolution,turchin2009secular,turchin2010political,turchin2016ages,acemoglu2013nations,gilens2014testing}. A major observation is that major political crises appear every few centuries and are rather common than an exception, while longer periods of ongoing political instability are also possible.

Current models of recurring societal instability are mainly macroscopic in nature \cite{turchin2009secular,turchin2010political,kondor2023explaining,wittmann2024demographic,roman2017coupled}, while the connection to microscopic/local interactions is still largely unclear. Here, network models \cite{newman2018networks,albert2002statistical,boccaletti2014structure} and opinion dynamic models \cite{clifford1973model,sakoda1949minidoka,schelling1971dynamic,axelrod1997dissemination,sznajd2000opinion,galam2002minority,deffuant2000mixing,hegselmann2015opinion,castellano2009statistical,starnini2025opinion} might come into play as they provide possible mechanisms for how macroscopic behavior can emerge from local interactions. Adaptive network models \cite{gross2008adaptive,rosvall2003modeling,rosvall2006modeling,holme2006nonequilibrium,kimura2008coevolutionary,auer2015dynamics,min2025aging,durrett2012graph,djurdjevac2024co,goto2024onset,gonzalez2025evidence} combine both and will serve us as a model paradigm in the following. 

In particular, we will study a minimal network model where the political power of an elite is represented as directed links, while its political orientation is represented as a node color. The model will be based on two main assumptions: the cumulative advantage effect of political power as well as on intra-elite conflict.

The cumulative advantage effect \cite{yule1925ii,simon1955class,price1976general,barabasi1999emergence} has been formulated with different names and also in terms of political power \cite{rigney2010matthew,perc2014matthew}. A recent general model uses only local rules in a redirection mechanism \cite{krapivsky2017emergent,krapivsky2024magic}, which we will use here as a basis. We define directed links as units of political power, and one can assume that the higher the political office, the more power an individual has. Furthermore, one can expect to have different magnitudes of power in a broad distribution, which also emerges from the local redirection mechanism \cite{krapivsky2017emergent,krapivsky2024magic}. 

Intra-elite conflict is hypothesized to be one of the key drivers of recurring political instability by Turchin and colleagues \cite{goldstone1991revolution,turchin2009secular,turchin2010political}. In order to model intra-elite conflict, each agent has a fundamental political orientation. We will define power as the ability to influence others, in particular, to replace opponents with peers of the same political idea. While a political office gives its holder the power to fulfill the intended tasks, it can also be used for personal benefits, such as to marginalize opponents.

\section{Model}
\begin{figure}[htbp]
\includegraphics[width=1\linewidth]{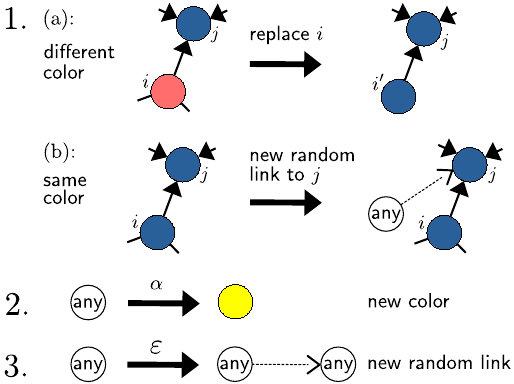}
\caption{Schematic model description} 
\label{fig:model_scheme} 
\end{figure} 
Using these two main assumptions, let us introduce a model with $N$ agents of the upper class of a society. Each agent $i$ has a fundamental political view $\sigma_i$, which is represented by a natural number and visualized by a color without intrinsic difference between different views. Power dependencies are modeled by directed links between agents. A link from agent $i$ to $j$ states that $i$ depends on $j$, or in other words $j$ has a unit of power on $i$. A schematic description of the model algorithm is shown in Fig. \ref{fig:model_scheme}. In detail, the algorithm reads:
\begin{enumerate}
    \item Choose a random agent $i$. If there is at least one outgoing link (dependency), randomly choose one outgoing neighbor $j$ and compare the mutual states $\sigma_i$ and $\sigma_j$ of the two nodes: 
    \begin{enumerate}
        \item If $\sigma_i \neq \sigma_j$, agent $j$ uses its power to remove $i$ including all its links and introduces a new agent $i'$ of the same color as $j$ and a link pointing to $j$.
        \item Else if $\sigma_i = \sigma_j$, another existing random agent is chosen that connects with a link pointing to $j$.
    \end{enumerate}
    \item With probability $\alpha$, a random existing agent changes its color to a new color that did not exist yet.
    \item With probability $\varepsilon$, a new link is added between two existing random agents.
\end{enumerate}
Each of the three steps are performed $N$ times, which defines one time unit $t \rightarrow t + 1$. Initially, the network starts with a circle of the same color, with each node having one in-link and one out-link. This ensures that all agents have the same starting conditions and that in the case of $\varepsilon=0$ the algorithm cases (a) and (b) are used. However, initial conditions were found to be irrelevant if the parameters are not zero.

If agents $j$ have different views than $i$, intra-elite conflict causes destruction, e.g. an agent can marginalize opponents with differing views. However, political unity is used to increase power. 

Probability $\alpha$ introduces a random influx of new ideas as society generates new political ideas that can eventually grow into a dominant political landscape, but are most likely to be forgotten.

The probability $\varepsilon$ creates independent random  links. This results in constraints independent of hierarchies emerging from existing power, such as official hierarchies with leading politicians on top. It can also be thought of as constraints between different sectors, such as political, economic, scientific, cultural, or religious leaders. Note that it is also possible that much less powerful agents can replace powerful agents by having the right link, which can be thought of as the right information that can ruin a career. Even powerful agents with many in-links face constraints from others.

Note that the model does not impose standard opinion dynamics on the political orientations. We motivate this with the impressionable years hypothesis \cite{newcomb1943personality,krosnick1989aging} that the political view of an individual is formed in younger years and remains largely unchanged in its foundations during a life span. One may, however, interpret the replacement step 1(a) also as an opinion change under a simultaneous drop of all old (political) connections. This is mathematically identical to a replacement of agent $i$ with a different agent $i'$ and may perhaps be absolutely not unrealistic. 

\section{Results}
\begin{figure}[htbp]
\centering
\includegraphics[width=1\linewidth]{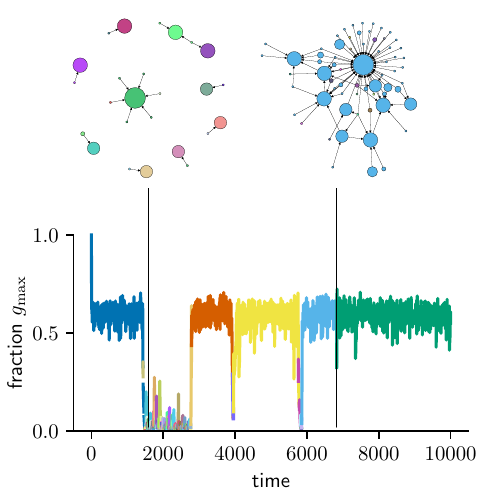}
\caption{Example time series of the dominant color fraction $g_\mathrm{max}$ for a system of $N=10^4$ agents, $\varepsilon=0.2$ and $\alpha=0.13$ close to the transition point $\alpha_\mathrm{c}$. Initially the network starts with a circle of the same color. The model exhibits punctuated equilibria with phases of persisting disunity and unity. Two example subnetworks at $t=1600$ and $t=6820$ contain the 10 nodes with the largest in-degree and a randomly chosen sample of 1\% (rounded down) of their in-degree neighbors. A few largest nodes are most influential for the model dynamics with each node trying to build its own star-like structure. Node sizes are logarithmically scaled by their in-degrees in the full network.}
\label{fig:example} 
\end{figure}

\begin{figure}[htbp]
\includegraphics[width=1\linewidth]{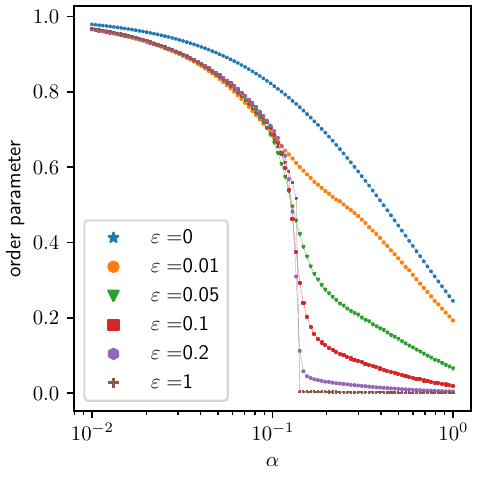}
\caption{The average order parameter for different parameter values for a system of $N=10^4$ agents and $10^6$ time units. The average is taken after $10^5$ time units. A phase transition appears for $\varepsilon \rightarrow 1$ and is suppressed for a fat tailed degree distribution for smaller $\varepsilon$.}
\label{fig:transition_eps}
\end{figure}

\begin{figure}[htbp]
\includegraphics[width=1\linewidth]{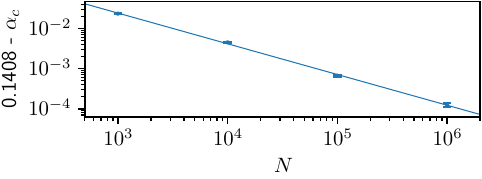}
\caption{Finite size scaling of the transition value $\alpha_\mathrm{c}$ for different system sizes $N$ for the phase transition for $\varepsilon=1$. $10^7$ time units were simulated for $N=10^3$, $10^6$ for $N=10^4$ and $10^5$ for larger $N$.}
\label{fig:transition_N}
\end{figure}

An example simulation is shown in Fig. \ref{fig:example}. The model exhibits punctuated equilibria close to the transition point $\alpha_\mathrm{c}$, which is further investigated in Figs. \ref{fig:transition_eps} and \ref{fig:transition_N}. For lower $\alpha$, quasi cycles appear with changes of the largest color and a characteristic period length. Stability within the network is maintained mainly by a few hubs in a core structure that create leaf nodes of the same color. Eventually, they are replaced, which opens the possibility for a new color to become dominant. For values of $\alpha$ close to the transition $\alpha_\mathrm{c}$, this may be interrupted by a longer phase of disunity without a color taking over the majority, as shown in Fig. \ref{fig:example}. For large $\alpha$, this behavior persists indefinitely. 

Star-like structures are emerging from the local preferential process along with a core structure of hub nodes, as discussed in \cite{krapivsky2017emergent}. Triadic closure only occurs by chance with probability $\varepsilon$, which effectively becomes irrelevant for this model, which can also be seen in the example as almost none of the random neighbor nodes are connected to each other. Every node is trying to build its own star structure, due to the model algorithm cases (a) and (b).  The total number of links is smaller in disunited phases, as nodes with different colors are being replaced, while also deleting all their links. However, if the colors are the same, random links that are created with probability $\varepsilon$ can be maintained and used to grow more links in the more united phases with a majority color. Note that agents are acting intrinsically the same, hub nodes only get more attention due to their larger in-degree.

One can define an order parameter $ B =(N g_\mathrm{max} - 1) / (N - 1)$ \cite{binder1981static}, with $g_\mathrm{max}$ being the fraction of the dominant color of agents within all nodes $N$ (majority color probability). The order parameter $B$ is 1 if only one color exists and 0 if every agent has a different color, while $g_\mathrm{max} \approx B$ has a minimum of $1/N$.

For different parameters, the order parameter is shown in Fig. \ref{fig:transition_eps}. In the limit $\varepsilon \rightarrow 0$, the dominant fraction is interrupted by new ideas flowing in at rate $\alpha$, resulting in $N$ picks with replacement
\begin{align*}
     B(\varepsilon=0) \approx \lim_{\varepsilon \rightarrow 0} g_\mathrm{max} = \left(1 - \frac{\alpha}{N}\right)^N \approx \exp(-\alpha)
\end{align*}
For this case, the model is dominated by a redirection process \cite{krapivsky2017emergent,krapivsky2024magic} with a single star structure emerging. The dominant color changes if the central hub changes color due to probability $\alpha$.

The probability density distribution of in-degrees $P(k)$ approaches a power law distribution for small $\varepsilon$, due to the multiplicative process \cite{newman2005power}. An analytical approximation for small $\alpha \rightarrow 0$ and $\varepsilon \rightarrow 0$ can be derived in analogy to \cite{zanette2020fat} in terms of a multiplier $\mu$ and a reset probability $q$. Resets are performed in the case of two agents with different colors being picked with probability $q = 1 - \exp (-2\alpha)$. For estimating the multiplier, one can count the number of new links being introduced during a time unit being
\begin{align}
    N (1 - q) \approx N(\mu - 1) \sum_{i=1}^N k_i^\mathrm{in} .
\end{align}
In-degrees $k_i^\mathrm{in}$ can be further approximated by its mean value $\langle k_i^\mathrm{in} \rangle$. Its value can be calculated by an equilibrium of rates $0 = N(1-q) - 2Nq\langle k_i^\mathrm{in} \rangle$. In analogy to \cite{zanette2020fat} this then leads to a power law exponent
\begin{align}
    \gamma = 1 + \log_\mu \left(\frac{1}{1 - q} \right) \approx 1 + \frac{2 \alpha}{\ln(3 - 2 \exp(-2\alpha))} .
\end{align}
In the limit of $\alpha \rightarrow 0$ and $\varepsilon \rightarrow 0$ the quantitative values align with simulation results of $ \lim_{(\alpha \rightarrow 0, \varepsilon \rightarrow 0)} \gamma \approx 1.5$. For larger $\alpha$ the distribution is becoming less broad. Note that random links are needed to create longer tree-like structures, while for $\varepsilon=0$ only a single star structure emerges.

In the limit $\varepsilon \rightarrow 1$ for a random network, a phase transition occurs at $\alpha_\mathrm{c} \approx 0.14$ as seen in Fig. \ref{fig:transition_eps}. Fig. \ref{fig:transition_N} further shows that the transition value $\alpha_\mathrm{c}$ changes with system size $N$. The phase transition for $\varepsilon \rightarrow 1$ can be understood with a mean field approach: colors can only spread with nodes being replaced in the case of $\sigma_i \neq \sigma_j$. If this process is omitted by new colors emerging with probability $\alpha$, the phase transition occurs. We can model it in terms of the dominant color fraction $g_\mathrm{max}$. The dominant color can grow in the case of different colors with probability $g_\mathrm{max} (1 - g_\mathrm{max})$, which uses a simplification that two nodes are picked with uniform probability. When two nodes are picked, the probability that one of them is replaced is indicated by $p_\mathrm{r}$. As $g_\mathrm{max}$ decreases due to new colors with probability $\alpha$, one can then formulate
\begin{align}
   \frac{\mathrm{d} g_\mathrm{max}}{\mathrm{d}t} = p_\mathrm{r} g_\mathrm{max} (1 - g_\mathrm{max}) - \alpha g_\mathrm{max}.
\end{align}
The steady state solution $\mathrm{d} g_\mathrm{max} / \mathrm{d}t = 0$ yields 
\begin{align}
    g_\mathrm{max} = 1 - \alpha / p_\mathrm{r},
\end{align} 
and a transition value $\alpha_\mathrm{c} \approx p_\mathrm{r}$, which qualitatively aligns with the simulation results.

In order to account for a possible early warning measure, one can count how many in-links a color has in relation to all links
\begin{align}
    \phi_\mathrm{in} =  \frac{\sum_{i=1}^N \delta_{\sigma_i \sigma_\mathrm{max}^\phi}k_i^\mathrm{in}}{\sum_{i=1}^N k_i^\mathrm{in}} \quad,
\end{align}
with $\delta_{\sigma_i \sigma_\mathrm{max}^\phi}$ being a Kronecker-Delta, which is 1 if $\sigma_i = \sigma_\mathrm{max}^\phi$ and 0 otherwise. $\sigma_\mathrm{max}^\phi$ resembles the color that gives the largest $\phi_\mathrm{in}$, which in most cases is the dominant color of the system. Similarly, the fraction of the dominant color could be expressed as $g_\mathrm{max} = \sum_{i=1}^N \delta_{\sigma_i \sigma_\mathrm{max}^g} /N$, in this case $\sigma_\mathrm{max}^g$ being the color that gives the largest $g_\mathrm{max}$, being the dominant color in the system. 

\begin{figure}
    \centering
    \includegraphics[width=1\linewidth]{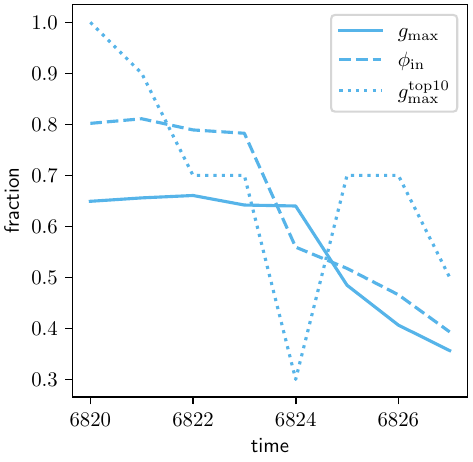}
    \caption{Example collapse of a dominant color $g_\mathrm{max}$, in-degree weighted color fraction $\phi_\mathrm{in}$ and the dominant color fraction of the 10 largest nodes by in-degree $g_\mathrm{max}^\mathrm{top10}$ for the simulation in Fig. \ref{fig:example}. The dominant fraction $g_\mathrm{max}$ lags behind $\phi_\mathrm{in}$ and partially behind $g_\mathrm{max}^\mathrm{top10}$. An example subnetwork for $t=6280$ can be observed in Fig. \ref{fig:example}.}
    \label{fig:correlation_example}
\end{figure}

An example temporal behavior of $\phi_\mathrm{in}$ and $g_\mathrm{max}$ can be observed in Fig. \ref{fig:correlation_example}. One can observe that $g_\mathrm{max}$ is lagging behind $\phi_\mathrm{in}$. Additionally, $g_\mathrm{max}^\mathrm{top10}$, being the dominant color fraction of the 10 largest nodes by in-degree, behaves in many time units similarly to $\phi_\mathrm{in}$, especially when changes are large. After the collapse, the previously dominant color gains back some advantage in terms of the 10 largest nodes. The next largest nodes, e.g. 11th to 20th, are likely to have the previously dominant color and take the places of the previously largest nodes, which are destroyed. Nodes with new colors need time to grow to the largest nodes in the system. Note that even if a destruction of the largest nodes does not result in a new dominant color, the majority of the dominant color can be more vulnerable in later time units.

\begin{figure}
    \centering
    \includegraphics[width=1\linewidth]{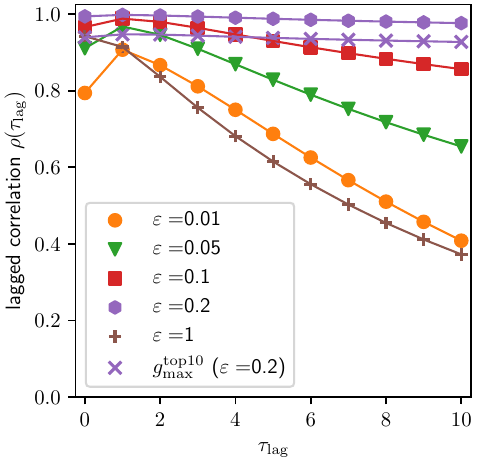}
    \caption{Lagged correlation $\rho(\tau_\mathrm{lag})$ of the dominant fraction $g_\mathrm{max}$ and $\phi_\mathrm{in}$ for different lagged time units $\tau_\mathrm{lag}$ for systems with $N=10^4$ and $\alpha=0.13$ close to the transition point $\alpha_\mathrm{c}$. Simulations with $10^6$ time units were used with correlations measured after $10^5$ time units. An increased correlation is observed for a lagged interval, which is also observed for just the dominant fraction of the largest 10 nodes by in-degree $g_\mathrm{max}^\mathrm{top10}$, with example subnetworks given in Fig. \ref{fig:example}.}
    \label{fig:correlation_statistics}
\end{figure}

Fig. \ref{fig:correlation_statistics} further quantifies the correlation in terms of a lagged Pearson correlation $\rho(\tau_\mathrm{lag})$ in dependence on a time lag $\tau_\mathrm{lag}$ in time units $t$. For $\varepsilon \rightarrow 0$, the correlation cannot be properly calculated, as only a single center node dominates the dynamics in a star-like structure, while the other nodes are simple leaf nodes, resulting in a constant $\phi_\mathrm{in}=1$. For $\varepsilon \rightarrow 1$, hubs do not dominate the dynamics as all nodes have large degrees with a continuously decreasing correlation for longer time lags. 

All examples in Fig. \ref{fig:correlation_statistics} show Granger-Causality with $p < 0.0001$ in causing $g_\mathrm{max}$, implying that $\phi_\mathrm{in}$ and $g_\mathrm{max}^\mathrm{top10}$ have predictive information. A huge benefit of $g_\mathrm{max}^\mathrm{top10}$ is that it only uses a few node states and does not need structural information, similar to the measures described in \cite{horstmeyer2020predicting}.

\section{Discussion}
We have studied a network model for power struggle which relies on purely local rules, and is based on the cumulative advantage effect of political power as well as intra-elite conflict. We showed how punctuated equilibria appear and how the local preferential attachment mechanism \cite{krapivsky2017emergent,krapivsky2024magic} generates broad degree distributions and a core network structure with hub nodes. 

The probability $\alpha$ of generating new colors is necessary to drive the system to new dominant colors, as otherwise only a single color would persist. It also controls the phase transition with punctuated equilibria appearing close to the transition point $\alpha_\mathrm{c}$. Instability phases between transitions of the dominant color can be very long or a transition can also appear with up to no instability phase in between, as shown in Fig. \ref{fig:example}. This aligns with historical observations \cite{goldstone1991revolution,turchin2009secular} where possible outcomes after major political crises can be another elite group taking over or a phase of political disunity with potentially civil unrest can persist.

The probability $\varepsilon$ of creating random links is crucial to create hierarchical trees of dependencies and, in general, a more complex network structure. We observe  a sharp phase transition for $\varepsilon=0$ which softens for $\varepsilon>0$, as can be seen in Figs.\ \ref{fig:transition_eps} and \ref{fig:transition_N}. This aligns well with the known behavior of the Ising model and the Potts model on scale-free graphs \cite{herrero2004ising,dorogovtsev2004potts} with hub nodes acting as pinning sites.

Empirical distributions of political power lack detailed quantitative support \cite{perc2014matthew}. In addition, misuse of political power and its effects are intrinsically invisible as they are useful to corrupt actors only while they remain hidden, known as the hidden cumulative advantage (Matthew) effect \cite{rigney2010matthew}. Proxy variables are useful for bypassing such problems, which are widely used by Turchin and colleagues. As one main result of the model, the dominant color of a few top nodes such as $g_\mathrm{max}^\mathrm{top10}$ has been found to represent the dynamics and can especially be used as predictive information for the temporal behavior. Although $\phi_\mathrm{in}$ is easy to calculate within this model, $g_\mathrm{max}^\mathrm{top10}$ is more accessible and potentially observable as a proxy in empirical data. 

Let us further discuss the model in the context of related research and w.r.t.\ possible extensions. As links reflect political power in the model, note that one could consider the in-degree distribution as an abstract representation of the distribution of power within a model society. This could potentially relate to the concept of elite overproduction \cite{turchin2009secular,turchin2010political,turchin2016ages}, which postulates that a growing number of elite aspirants are drivers of intra-elite conflict due to limited positions. Broader in-degree distributions have more agents with more power and its exponent could serve as a proxy for elite overproduction. While in the current version of the model the in-degree distribution remains relatively constant, an explicit mechanism for elite overproduction could be an interesting extension. 

As political power and wealth are naturally intertwined, another possible extension could include the influence of wealth as already studied separately \cite{jochim2025self}. Furthermore, this model could also provide a mechanism for social cohesion in growing groups \cite{fenoaltea2023phase} and its connection to civil unrest and scaling \cite{lee2020scaling,braha2024phase} should be further investigated.

Last, not least, the role the phase transition could play in a social context may also be explored. Note that there is the possibility to extend the model with a self-organized critical mechanism, for example by letting step 1.(b) of the model take care of creating new colors. We leave the discussion of the implications and possible realizations for future work. 

\bibliography{literature.bib}

\end{document}